\documentclass[fleqn,twoside]{article}
\usepackage{espcrc2}

\usepackage{epsfig}

\newif\ifpdf
\ifx\pdfoutput\undefined
\pdffalse % we are not running PDFLaTeX
\else
\pdfoutput=1 % we are running PDFLaTeX
\pdftrue
\fi

\newcommand{\lesssim}{\raisebox{-.25em}{$\stackrel{\mbox{\normalsize $<$}}{\sim}$}}

\newcommand{\bnP}{\bar {\cal P}}

\newcommand{\cP}{{\cal P}}
\newcommand{\DSppP}{{{D\!\!\!\!\hspace{0.04cm}\slash}}_c^\perp}
\newcommand{\DSppPl}{\stackrel{\leftarrow}{{{D\!\!\!\!\!\hspace{0.04cm}\slash}}}\!\!{}_c^\perp}
\newcommand{\DSppPr}{\stackrel{\rightarrow}{{{D\!\!\!\!\!\hspace{0.04cm}\slash}}}\!\!{}_c^\perp}
\newcommand{\bsigma}{\mbox{\boldmath $\sigma$}}

\newcommand{\bn}{{\bar n}}

\newcommand{\nslash}{n\!\!\!\slash}
\newcommand{\vslash}{v\!\!\!\slash}

\newcommand{\bnslash}{\bar n\!\!\!\slash}
\newcommand{\nn}{\nonumber}

\title{Factorization, Effective Field Theory, and $B\to D^{(*)}X$ Decays}

\author{Iain W. Stewart 
\address{Institute for Nuclear Theory, University of Washington, Seattle, WA 
98195} 
\thanks{INT-PUB 02-45 Invited plenary talk at the {\em 5th International 
Conference on Hyperons, Charm and Beauty Hadrons}, UBC, Vancouver, June 2002.}
}
       
\begin{document}

\begin{abstract}

In this proceedings I review the soft-collinear effective theory (SCET), an
effective theory for energetic particles.  I also discuss factorization in
exclusive and inclusive $B\to D^{(*)}X$ decays, and tests which can help
distinguish whether factorization is a result of a large energy limit, the large
$N_c$ limit, or a combination of the two.

\vspace{1pc}
\end{abstract}

\maketitle

\section{Introduction}

Effective field theory methods provide us with a useful tool for separating
short and long distance fluctuations. Examples include the Electroweak
Hamiltonian (separating $m_{b,c}\ll m_W$), Heavy Quark Effective Theory
(separating $\Lambda_{\rm QCD}\ll m_b$), and Chiral Perturbation Theory
(separating $p_\pi \ll \Lambda_{\chi}$). In this talk I discuss an effective
field theory that has been developed for processes with energetic hadrons,
referred to as the Soft-Collinear Effective Theory
(SCET)~\cite{bfl,bfps,cbis,bpssoft,bfprs} (separating $\Lambda_{\rm QCD}\ll Q$).

For high energy processes the separation of long $p\sim \Lambda_{\rm QCD}$ and
short $p\sim Q$ distance QCD fluctuations is often referred to as
factorization. Familiar examples include Deep Inelastic Scattering, Drell Yan,
and exclusive form factors~\cite{review}. Factorization is also relevant to
B-decays to light hadrons due to the large energy released. Examples of such
processes include the decays $B\to D\pi$, $B\to \pi\pi$, $B\to K\pi$, $B\to
K^*\gamma$, the large recoil region in $B\to\pi\ell\nu$, $B\to \rho\ell \nu$,
and $B\to K\ell^+\ell^-$, and the endpoint spectra of the inclusive decays $B\to
X_u \ell\nu$ and $B\to X_s\gamma$. In particular there has been much discussion
of the nature of factorization for the non-leptonic decays $B\to \pi\pi$ and
$B\to K\pi$ which are relevant to measuring the CKM angles $\alpha$ and
$\gamma$~\cite{Bpipi}.  A possible source of confusion is that in B-physics the
word ``factorization'' is also used to refer to the process of approximating the
matrix element of a four quark operator by the product of two non-interacting
bilinear currents (the latter definition is often used even in cases where this
is not a justified procedure). To distinguish the two meanings I will refer to
the latter as ``4q-factorization''.

The idea behind SCET is to provide a systematic framework for studying inclusive
and exclusive processes with energetic particles.  This includes the
investigation of power corrections and parameterization of non-perturbative
effects by matrix elements of operators. To review the status of SCET I have
broken the discussion into several categories: fields and power counting, gauge
symmetries, spin symmetries, reparameterization invariance, Sudakov logarithms,
and factorization.  As an application I discuss the proof of factorization for
$B\to D\pi$, $B\to D^*\pi$, $B\to D\rho$, $B\to D^*\rho$, and similar decays in
the large energy limit $Q=\{m_b,m_c,m_b-m_c\}\gg \Lambda_{\rm QCD}$.  In this
case the separation of short and long distance scales leads to a type of
4q-factorization which has well defined scheme and scale dependence. I also
discuss how the large $N_c$ limit leads to 4q-factorization, and what
measurements can be used to distinguish the importance of these two limits of
QCD for $B\to D^{(*)}X$ decays.

\section{Soft-Collinear Effective Theory}

\begin{table*}[htb]
\caption{Power counting for momenta, SCET fields, label operators
($\bnP,\cP_\perp^\mu$, $\cP^\mu$) and Wilson lines ($W$, $S_n$, $Y_n$).  }
\hspace{0.2cm}
\label{table_pc}
\begin{center}
\begin{tabular}{cl|c|clc}
\hline
  Type & Momenta $(+,-,\perp)$\hspace{0.4cm} 
   & \hspace{0.2cm}Field Scaling  \hspace{0.cm} 
   & \hspace{0.2cm}Operators\hspace{0.2cm} \\ 
   \hline
  collinear & $p^\mu\sim (\lambda^2,1,\lambda)$ \hspace{0.2cm}
   & \hspace{0.2cm} $\xi_{n,p}\sim \lambda$ & $\bnP$, $W_n$ $\sim\lambda^0$ \\
  && ($A_{n,p}^+$, $A_{n,p}^-$, $A_{n,p}^\perp$) $\sim$ 
  ($\lambda^2$,$1$,$\lambda$) & \hspace{0.4cm} $\cP_\perp^\mu\sim \lambda$ \\
   \hline
  soft &  $p^\mu\sim (\lambda,\lambda,\lambda)$ \hspace{0.25cm}
   & \hspace{0.8cm} $q_{s,p}\sim \lambda^{3/2}$ & \hspace{0.5cm} 
   $S_n \sim \lambda^0$ \\
  & & \hspace{0.2cm} $A_{s,p}^\mu\sim \lambda$ & \hspace{0.3cm} 
    $\cP^\mu\sim \lambda$ \\ \hline 
  usoft &  $k^\mu\sim (\lambda^2,\lambda^2,\lambda^2)$
   & \hspace{0.5cm} $q_{us}\sim \lambda^3$ & \hspace{0.4cm} $Y_n\sim\lambda^0$ \\
  & & \hspace{0.4cm} $A_{us}^\mu\sim \lambda^2$  \\
\hline
\end{tabular}
\end{center}
\end{table*}

In this section I give a brief overview of some of the important features of
SCET. For further details on the effective theory I refer the reader to the
literature~\cite{bfl,bfps,cbis,bpssoft,bfprs}. 

%%%%%%%%%%%%%%%%%%%%%%%%%%%%%%%%%%%%%%%%%%%%%%%%%%%%%%%%

$\bullet$ Fields and Power Counting. The effective theory is derived from QCD by
integrating out fluctuations with $p^2\gg Q^2\lambda^2$, where typically
$\lambda=(\Lambda_{\rm QCD}/Q)^k$ with $k=1$ or $k=1/2$.  Long distance
fluctuations with $p^2\,\lesssim\, Q^2\lambda^2$ are responsible for the
infrared divergences and are described by effective theory fields.  Typical
processes involve particles with collinear momenta and either soft or usoft
momenta. In Table~\ref{table_pc} the scaling of these momenta with the expansion
parameter $\lambda$ are given, together with the quark and gluon fields assigned
to each type of fluctuation.

The momentum scales $Q$, $Q\lambda$, and $Q\lambda^2$ can be separated by making
phase redefinitions to pull out the larger momenta, $\phi_n(x) =\sum_p
e^{-ip\cdot x} \phi_{n,p}(x)$. (An interesting alternative working purely in
position space can be found in Ref.~\cite{BCDF}.)  Derivatives on the new fields
then give $\partial^\mu\phi_{n,p}(x)\sim (Q\lambda^2)\, \phi_{n,p}(x)$, while
larger momenta are picked out by introducing the label operators~\cite{cbis}
$\bnP, \cP^\mu$, so for example $\bnP\, \xi_{n,p}=(\bn\cdot p)\,\xi_{n,p}$.

Integrating out the offshell fluctuations also builds up Wilson lines in
effective theory operators.  These include a Wilson line $W[\bn\cdot A_{n,q}]$
which is build out of collinear gluons fields that are $\sim \lambda^0$ in the
power counting. An example where this appears is the matching of full QCD
heavy-to-light current $\bar u \Gamma b$ onto the SCET current~\cite{bfps},
$\bar\xi_{n,p}\,W\,\Gamma\, h_v$. In fact the $\bn\cdot\! A_{n,q}$ field can be
traded for the Wilson line since the covariant derivative $i\bn\cdot
D_c=\bnP+g\bn\cdot A_{n,q} = W\, \bnP\, W^\dagger$.  Soft Wilson lines
$S_n[n\cdot A_s]$ are also built up by integrating out offshell
fluctuations~\cite{bpssoft}.

A gauge invariant power counting for fields can be fixed by demanding that
the kinetic terms in the action are order $\lambda^0$~\cite{bfl,bfps}. The power
counting for an arbitrary diagram, $\lambda^\delta$, can then be determined
entirely from its operators using~\cite{bpspc}
\begin{eqnarray}
  \delta = 4 \!+\! \sum_k (k\!-\!4) [V_k^C \!+\! V_k^S \!+\! V_k^{SC}] 
     \!+\! (k\!-\!8) V_k^{U}\,. \nn
\end{eqnarray}
Here $V_k^{C,S,SC,U}$ count the number of order $\lambda^k$ operators which have
collinear fields, soft fields, both, or neither respectively. Since $\bnP\sim
\lambda^0$ in the power counting the Wilson coefficients $C(\bnP)$ are arbitrary
functions of this operator~\cite{cbis}, and can be determined by matching. More
generally we have functions $C(\omega_i) \prod_i \delta(\omega_i-\bnP)$ where
the delta functions are inserted inside collinear operators in the most general
locations allowed by gauge and reparameterization invariance.

%%%%%%%%%%%%%%%%%%%%%%%%%%%%%%%%%%%%%%%%%%%%%%%%%%%%%%%%

$\bullet$ Gauge Symmetries: The structure of operators containing factors of $W$
or $S_n$ is protected by collinear, soft, and usoft gauge
transformations~\cite{cbis,bpssoft}. These are gauge transformations $U(x)$,
where $\partial^\mu U(x)$ scales like a collinear, soft, or usoft momentum. A
simple example is the leading order heavy-to-light current with an usoft 
heavy quark $h_v$,
\begin{eqnarray}
 J_0 &=& C_i(\bnP)\: \bar\xi_{n,p}\, W\, \Gamma_i\, h_v \,.
\end{eqnarray}
Under an usoft gauge transformation $U_u$, all the fields transform and we have
$J_0\to C_i(\bnP)\: \bar\xi_{n,p} U_u^\dagger\, U_u W U^\dagger_u \,
\Gamma_i\,U_u h_v=J_0$.  Under a collinear gauge transformation $U_c$ the field
$h_v$ does not transform and we have $J_0\to C_i(\bnP)\: \bar\xi_{n,p}
U_c^\dagger\, U_c W\, \Gamma_i\, h_v=J_0$. Since $\bnP$ does not commute with
$U_c$ the function $C_i(\bnP)$ can only act on the product
$\bar\xi_{n,p}W$. Another example of a gauge invariant operator is the leading
order collinear quark Lagrangian~\cite{bfps,cbis}
\begin{eqnarray}\label{Lc0}
{\cal L}_0 = \bar \xi_{n,p'} \left\{n\cdot iD \!+\!
 i\DSppP \frac{1}{i\bn\!\cdot D_c} i\DSppP \right\} 
 \frac{\bnslash}{2}\: \xi_{n,p} ,
\end{eqnarray}
where $n\cdot D= in\cdot\partial + gn\cdot A_{us} + gn\cdot A_{n,q}$ contains
both usoft and collinear gauge fields, while $iD_c^{\perp\,\mu}=\cP_\perp^\mu +
gA_{n,q}^{\perp\,\mu}$ and $i\bn\cdot D_c = \bnP+g\bn\cdot A_{n,q}$ are purely
collinear.  For the explicit form of the leading order gluon action we refer the
reader to Ref.~\cite{bpssoft}. For higher order terms in the collinear quark
action we refer the reader to Refs.~\cite{chay1,mmps,chay2}, and for terms in
the mixed collinear-usoft quark action to Ref.~\cite{BCDF}
\begin{figure}[t!]
\begin{center}
 \includegraphics[height=1.in]{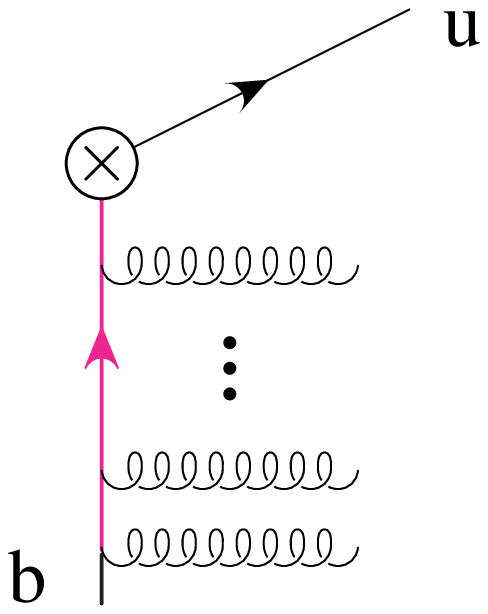}
 \hspace{0.5cm} \raisebox{1.4cm}{\Large $\Longrightarrow$} \hspace{0.5cm}
 \includegraphics[height=1.in]{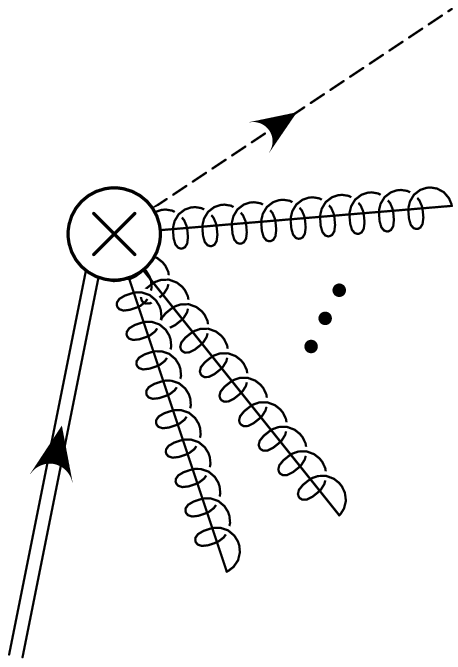}
\caption{Tree level matching for the leading order heavy-to-light current. The
gluon fields shown here are $\bn\cdot A_{n,q}$ which are order $\lambda^0$.}
\label{fig:match}\vspace{-0.7cm}
\end{center}
\end{figure}
 
%%%%%%%%%%%%%%%%%%%%%%%%%%%%%%%%%%%%%%%%%%%%%%%%%%%%%%%%

$\bullet$ Reparameterization Invariance: In the heavy quark effective theory
Lorentz invariance is broken by the velocity vector $v^\mu$, with $v^2=1$, which
labels the fields, $h_v$. However this symmetry is restored order by order in
the power counting by a reparameterization invariance (RPI) under small changes
in $v^\mu$~\cite{LM}. A similar situation arises in SCET where collinear fields
are defined by introducing two auxiliary light-like vectors, $n$ and $\bn$, such
that $n\cdot \bn=2$. For SCET the study of RPI was initiated in
Ref.~\cite{chay1} and generalized to the three most general classes of allowed
transformations in Ref.~\cite{mmps}:
\begin{eqnarray}
\label{repinv}
\mbox{(I)} \left\{ \begin{tabular}{l}
  $n_\mu \to n_\mu + \Delta_\mu^\perp$ \\
  $\bn_\mu \to \bn_\mu$
\end{tabular} \right.\quad
\mbox{(II)} \left\{ \begin{tabular}{l}
  $n_\mu \to n_\mu$ \\
  $\bn_\mu \to \bn_\mu + \varepsilon_\mu^\perp$
 \end{tabular} \right.\nn \\
%\begin{center}
\mbox{(III)} 
\left\{
\begin{tabular}{l}
$n_\mu \to (1+\alpha)\, n_\mu$ \\
$\bn_\mu \to (1-\alpha)\, \bn_\mu$
\end{tabular}
\right. \,,\hspace{3.3cm} \nn
\end{eqnarray}
where $\alpha\sim \lambda^0$, $\epsilon^\perp\sim \lambda^0$, and
$\Delta^\perp\sim \lambda$. The transformations $(\rm I,II,III)$ constrain the
allowed form of collinear operators both within and between different orders in
$\lambda$. For instance they rule out additional operators such
as~\cite{mmps}
\begin{eqnarray}
 \bar\xi_{n,p'} \Big\{ i D_c^{\perp\,\mu}\: \frac{1}{i\bn\cdot D_c}\: 
  i D_{c\,\mu}^{\perp} \Big\} \frac{\bnslash}{2}\xi_{n,p}
\end{eqnarray}
in the Lagrangian in Eq.~(\ref{Lc0}), and fix the value of certain Wilson
coefficients in subleading terms in the collinear quark
Lagrangians~\cite{chay1,mmps}. The type-I transformations relate the Wilson
coefficients $B_i$ of the ${\cal O}(\lambda)$ suppressed heavy-to-light
currents~\cite{chay1}
\begin{eqnarray} \label{J1a}
 J_{1a} &=& B_i(\bnP)\: \bar\xi_{n,p'} \frac{\bnslash}{2} i\!\!\DSppPl W 
  \frac{1}{\bnP}\, \Gamma_i\: h_v,
\end{eqnarray}
to the coefficients $C_i(\bnP)$ of the leading currents. However, there are a
set of ${\cal O}(\lambda)$ suppressed heavy-to-light currents derived in
Ref.~\cite{bpspc,BCDF}
\begin{eqnarray}  \label{J1b}
 J_{1b} &=& E_i(\bnP)\: \bar\xi_{n,p'} \Gamma_i\:i\!\!\DSppPr W 
  \frac{1}{m_b}\,\frac{\nslash}{2}  h_v\,,
\end{eqnarray}
whose Wilson coefficients are not related to the $C_i(\bnP)$ under type I RPI
transformations. The set of heavy-to-light currents up to order $\lambda^2$
which have non-zero tree level matching have been derived in Ref.~\cite{BCDF}.

%%%%%%%%%%%%%%%%%%%%%%%%%%%%%%%%%%%%%%%%%%%%%%%%%%%%%%%%%

$\bullet$ Spin Symmetries: The spinors associated with collinear light quarks or
(u)soft heavy quarks have only two components. When 4-component Dirac spinors
are used this is embodied in the projector relations $P_n \xi_n =\xi_n$ and $P_v
h_v=h_v$, where $P_n = \nslash\bnslash/4$ and $P_v=(1+\vslash)/2$. For heavy
quarks $P_v$ projects out the anti-particle components and the leading
Lagrangian for the quarks, ${\cal L}_h =\bar h_v iv\cdot D h_v$ has an SU(2)
spin-symmetry~\cite{IW}.  For collinear quarks the projector $P_n$ eliminates
the two components of the spinor corresponding to motion in the direction
opposite to $n^\mu$. If the four component spinor is written
\begin{eqnarray} \label{spinor}
  u = \frac{1}{\sqrt{2}} \left( \begin{array}{c} \varphi \\ \frac{\bsigma\cdot
  {\bf p}}{p^0} \varphi \end{array} \right)\,,\quad
    \varphi=\left(\begin{array}{c} 1\\ 0 \end{array}\right) \mbox{ or }
    \left(\begin{array}{c} 0\\ 1 \end{array}\right)\,,\nn
\end{eqnarray}
then for motion in the $z$ direction the spinors $u_n^\dagger =
(\varphi^\dagger\: \varphi^\dagger \sigma_3)/\sqrt{2}$ are kept and the spinors
$u_\bn^\dagger =(\varphi^\dagger\, -\!\varphi^\dagger \sigma_3)/\sqrt{2}$ are
eliminated. The leading order Lagrangian is shown in Eq.~(\ref{Lc0}) and still
has both particles and antiparticles. It also has a helicity spin-symmetry with
generator $h=i\epsilon_\perp^{\mu\nu}[\gamma_\mu, \gamma_\nu]/4$. In
Refs.~\cite{DG,Charles} the simpler Lagrangian ${\cal L}_{\rm LEET} = \bar\xi_n
(\bnslash/2) in\!\cdot\! D\, \xi_n$ was discussed and observed to have a larger
SU(2) spin-symmetry. Unfortunately, this Lagrangian does not correctly describe
the dynamics of fast moving particles. For the correct Lagrangian in
Eq.~(\ref{Lc0}) the $\gamma_\perp^\mu \gamma_\perp^\nu\bnslash$ structure in the
second term leaves only the helicity generator of the SU(2)
unbroken~\cite{bfps,BH}. This becomes obvious when we write Eq.~(\ref{Lc0}) in
terms of a two-component field $\varphi_{n,p}$ (whose spinors are the two
$\varphi$'s)
\begin{eqnarray} 
 {\cal L}_0 &=& \varphi^\dagger_{n,p'} \Big\{ n\!\cdot\! iD 
  + iD_c^{\perp\mu} \frac{1}{i\bn\!\cdot\! D_c} i D_c^{\perp\nu} \\
 && \qquad \times (g^\perp_{\mu\nu}+ i \epsilon^\perp_{\mu\nu} \sigma_3)\Big\} 
  \varphi_{n,p}\,,\nn
\end{eqnarray}
where $g^\perp_{\mu\nu}=g_{\mu\nu} - \bn^{\mu}n^\nu/2 - n^\mu\bn^\nu/2$ and
$2\epsilon_\perp^{\mu\nu} = \epsilon^{\mu\nu\alpha\beta} \bn_\alpha n_\beta $.

The reduction in spin structures for the heavy and collinear fields have
observable consequences. In particular for the heavy-to-light currents $J_0$
only four spin structures $\Gamma_i$ are allowed (which can be chosen as $1$,
$\gamma_5$, and $\gamma_\perp^\mu$)~\cite{bfps}.  This reduces the number of
soft form factors for $B$ decays to a pseudoscalar meson from three to one
($f_+, f_0, f_T\to \zeta$), and the number for $B$ decays to a vector meson from
seven to two ($A_{0,1,2}, T_{1,2,3}, V\to \zeta_\perp,\zeta_\parallel)$. These
soft form factor relations are identical to those derived in Ref.~\cite{Charles}
using ${\cal L}_{\rm LEET}$, despite the fact that SCET does not have the SU(2)
spin-symmetry. This is because the relations are a consequence of projecting out
the suppressed $u_\bn$ components rather than from a residual spin-symmetry in
SCET. Hard contributions which violate these form factor relations were derived 
in Ref.~\cite{BF}.

%%%%%%%%%%%%%%%%%%%%%%%%%%%%%%%%%%%%%%%%%%%%%%%%%%%%%%%%

$\bullet$ Sudakov Logarithms: The SCET can be used to sum double Sudakov
logarithms~\cite{bfl}. For the inclusive decay $B\to X_s\gamma$ in the endpoint
region $m_B\!-\!E_\gamma/2\,\lesssim \Lambda_{\rm QCD}$ the scales are $Q=m_b$,
$Q\lambda=\sqrt{m_b\Lambda_{\rm QCD}}$, and $Q\lambda^2=\Lambda_{\rm QCD}$. The
logarithms can be summed in two steps by i) running in SCET from $Q$ to
$Q\lambda$, ii) running in a purely usoft theory from $Q\lambda$ to a scale of
order $Q\lambda^2$~\cite{bfl}. For $B\to X_s\gamma$ this usoft theory is just
like HQET with additional non-local operators. Since $(Q\lambda)^2$ is the
offshellness of the collinear particles and the interactions
factor~\cite{ks,bpssoft} these fields are integrated out before the second stage
of running. In the first stage of running the generic LO anomalous dimension for
double logarithms looks like~\cite{bfps}
\begin{eqnarray}
  \mu \frac{\partial}{\partial\mu} C(\mu,\bnP) = a\: \alpha_s(\mu) 
  \ln\Big(\frac{\mu}{\bnP}\Big)  C(\mu,\bnP) \,,
\end{eqnarray}
where $a$ is a number. The solution exponentiates due to the homogeneous nature
of the renormalization group equation. For this stage the running for $B\to
X_s\gamma$ and $B\to X_u \ell\nu$ differs.  In the second stage of running the
operators obtained by integrating out the collinear fields are $O(y)$ where the
coordinate $y$ is related to the dynamic nature of the matching scale. The
anomalous dimension for their Wilson coefficients have the structure~\cite{bfl}
\begin{eqnarray}
  \mu \frac{\partial}{\partial\mu} c(y,\mu) = \int dy'\: \gamma(y,y',\mu) \:
     c(y',\mu) \,.
\end{eqnarray}
With a proper choice of matching scales the running for $B\to X_s\gamma$ and
$B\to X_u \ell\nu$ is the same for this second stage. A similar procedure has
been applied to sum logarithms in the Upsilon decays $\Upsilon\to X\gamma$ in
the color octet channel~\cite{BCFLL}.

%%%%%%%%%%%%%%%%%%%%%%%%%%%%%%%%%%%%%%%%%%%%%%%%%%%%%%%%

$\bullet$ Factorization: In general proofs of factorization are simplified in
the effective theory by the nature of the operator formulation, which makes the
steps explicitly gauge invariant, and handles many classes of diagrams
simultaneously. The factorization between hard and collinear fluctuations or
soft and collinear fluctuations is simplified by the fact that it takes place at
the matching level, and is constrained by SCET symmetries.  The factorization
between collinear and usoft interactions occurs in the leading effective
Lagrangian and is simplified by the fact that certain cancellations are
therefore universal.  For instance, at lowest order the actions for usoft and
collinear particles can be factorized by a simple field redefinition on the
collinear fields~\cite{bpssoft}, $\xi_{n,p}=Y_n \xi_{n,p}^{(0)}$ and $A_{n,p} =
Y_n A_{n,p}^{(0)} Y_n^\dagger$, where $Y_n = P\exp [ig \int_{-\infty}^x\!\! ds\,
n\!\cdot\!  A_{us}(s n)]$. This transformation moves all leading order usoft
interactions out of the collinear Lagrangian ${\cal L}^{(0)}_c$ and into the
external operators and currents, after which cancellations due to the identity
$Y_n^\dagger Y_n=1$ for the usoft Wilson line, are more readily seen .

\section{$B\to D\pi$ factorization for $Q\gg\Lambda_{\rm QCD}$}

For decays like $B\to D\pi$ the energy of the outgoing meson in the rest frame
of the $B$ is large. It is therefore useful to consider this decay in the large
$Q$ limit where $Q=m_b$, $m_c$, or $E_\pi$. In this limit, SCET has been used to
prove the following factorization theorem~\cite{bps}
\begin{eqnarray} \label{fact1}
&& \langle D \pi | H_W  | B\rangle \\
&& = N\:  \xi^{\rm IW}(w_0)
 \int_0^1 dy\ { T(y,Q)}\ 
 {\phi_\pi(y)} + \ldots \,, \nn
\end{eqnarray}
where the ellipses denote terms that vanish as a higher power as $(\Lambda_{\rm
QCD}/Q) \to 0$.
\begin{figure}[t!]
\begin{center}
  \includegraphics[height=1.5in]{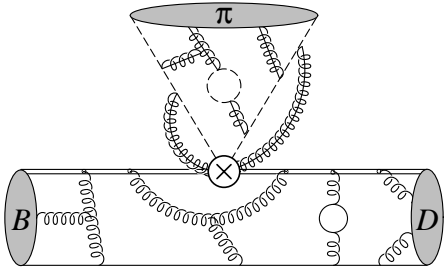}
 \vspace{-0.5cm}
\caption{How the factorization of interactions occurs in $B\to D\pi$. Gluons
with a line through them are collinear, while those without are soft or
usoft. \vspace{-0.8cm}}
\label{fig:bdpi}
\end{center}
\end{figure}
Here the electroweak Hamiltonian is $H_W = C_1 O_1 + C_8 O_8$, where $O_{1}=\bar
d_L \gamma^\mu u_L\bar c_L \gamma_\mu b_L$ and $O_{8}= \bar d_L T^A\gamma^\mu
u_L \bar c_L T^A\gamma_\mu b_L$.  Eq.~(\ref{fact1}) was proposed in
Ref.~\cite{PW}, proven to two-loops in Ref.~\cite{bbns}, and proven to all
orders in $\alpha_s$ in Ref.~\cite{bps}.  The idea behind the proof is shown in
Fig.~\ref{fig:bdpi}. Integrating out offshell fluctuations gives rise to SCET
operators $Q_{\{0,8\}} = [\bar\xi_{n,p'}^{(d)} W C_{0,8}(\bnP_\pm) \Gamma_\ell
\{1,T^A\} W^\dagger \xi_{n,p}^{(u)}] [\bar h_{v'}^{(c)} S \{1,T^A\}$ $\Gamma_h
S^\dagger h_v^{(b)}]$. The matrix element factors at lowest order because it is
possible to isolate all soft gluons in the $(\bar c b)$ bilinear and collinear
gluons in the $(\bar d u)$ bilinear and use the fact that the pion state is
purely collinear and the $B$ and $D$ are purely soft. Soft gluons are exchanged
between quarks in the $B$ and $D$, and the matrix element $\langle D^{(*)}| \bar
h_v \Gamma_h h_v | B\rangle$ gives the Isgur-Wise function $\xi^{\rm IW}$ in
Eq.~(\ref{fact1}). Collinear gluons build up the pion state and the matrix
element $\langle \pi | \bar\xi_n W \delta(\omega-\bnP_+)\Gamma_\ell W^\dagger
\xi_n |0\rangle$ gives rise to $\phi_\pi(x)$. Finally the hard kernel is related
to the Wilson coefficients $C_0(\bnP_+,\bnP_-)$ by $T(y) =
C_0((4y-2)E_\pi,2E_\pi)$.

The large $Q$ factorization theorem applies to all decays that have
contributions from the tree topology in Fig.~\ref{fig:bdpi} and are similar
kinematically to $B\to D\pi$. The list therefore includes $\bar B^0\to D^+\pi^-$
and $B^-\to D^0\pi^-$. It also includes the analogous $B\to D^{*}\pi$, $B\to
D\rho$, and $B\to D^*\rho$ modes.  For the $\bar B^0$ decays the predictions
agree well with factorization. However the predictions for $B^-$ decays receive
large corrections.  Using the recent CLEO measurements~\cite{cleoBDpi} gives
\begin{eqnarray}
  R = \frac{Br(B^-\to D^0\pi^-)}{Br(\bar B^0\to D^+\pi^-)} = 1.85\pm 0.13\,,
\end{eqnarray}
which corresponds to $30$--$40\%$ corrections for the $B^-$ matrix elements to
the $Q\to\infty$ prediction of $R\simeq 1$. In addition using the $Br(\bar
B^0\to D^0\pi^0)$ one can extract a non-zero strong phase difference between
isospin amplitudes for the $B\to D\pi$ modes~\cite{PN}, $\Delta\delta \simeq
27\pm 7^{\circ}$. Both of these effects are numerically of the size of
$\Lambda_{\rm QCD}/m_c$ corrections to Eq.~(\ref{fact1}). A parameterization of
$1/m_c$ corrections is possible using SCET, however they have not yet been
worked out.

\section{Large $Q$ or Large $N_c$ in $B\to D^{(*)}X$}

\begin{figure}[t!]
\begin{center}
  \includegraphics[height=1.8in]{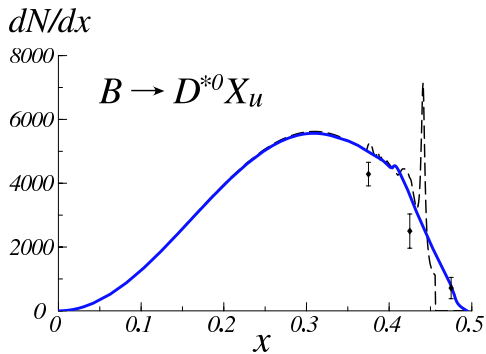}
\vspace{-0.5cm}
\caption{A measurement of the inclusive $m_X^2$ spectrum for $B\to D^{*0}X_u$
will give a direct handle on the region of validity of 4q-factorization. Here 
$x=|\vec p_D|$ and so large $x$ is small $m_X^2$. The
solid curve shows the prediction made using the combined large $N_c$ and SV
limits and boosted to the rest frame of the Upsilon(4S) (dashed is
unboosted)~\cite{bgps}.  The three data points from CLEO~\cite{CLEODX} are shown
that are not contaminated by charmed states in $X_u$.\vspace{-0.9cm}}
\label{fig:DXu}
\end{center}
\end{figure}

Taking the large $N_c$ limit of non-leptonic matrix elements of four quark
operators such as those in $H_W$ also leads to a type of 4q-factorization, but
without calculable perturbative corrections. In the large $N_c$ limit the 
predictions for a transition $B\to D^{(*)}X$ take the form
\begin{eqnarray} \label{fact2}
&& \hspace{-0.4cm}
\langle X D^{(*)} | (\bar c b)_{V-A}(\bar d u)_{V-A} | \bar B \rangle \\
 && \hspace{-0.4cm}
= \sum_{X',X''} \langle D^{(*)}X' | (\bar c b)_{V-A} | B \rangle 
 \langle X''| (\bar d u)_{V-A} | 0 \rangle\,,\nn
\end{eqnarray}
where $X=X'+X''$. For $B\to D\pi$ decays the leading order predictions from
large $N_c$ are numerically quite similar to those from large $Q$, however there
are many decays where this is not the case.  For instance, as the invariant mass
of the light meson state, $m_X$, increases large $Q$ factorization is expected
to break down, unlike large $N_c$ 4q-factorization~\cite{DG}.  In
Ref.~\cite{LLW} it was pointed out that the $m_X^2$ spectrum in $\bar B^0\to
D^*\omega\pi^-$ and $\bar B^0\to D^{*0}\pi^+\pi^+\pi^-\pi^-$ can be used to test
4q-factorization and the data agrees well out to $m_X^2\sim m_\tau^2$. This
test relies on numerically small contributions from terms in Eq.~(\ref{fact2})
with $X'\ne 0$. In Ref.~\cite{bgps} it was pointed out that this could be
remedied by taking the combined large $N_c$ and SV~\cite{SV} ($m_b\gg m_b-m_c
\gg \Lambda_{\rm QCD}$) limits, where SV suppresses the $X'$ contributions. An
additional test involving the inclusive $m_X^2$ spectrum for $B\to D^{*0}X_u$
was then proposed as shown in Fig.~\ref{fig:DXu}.  A measurement of the full
spectrum would test factorization all the way out to $m_X^2\sim 10\,{\rm
GeV^2}$.  Additional tests of 4q-factorization are also possible using exclusive
final states~\cite{bgps}. Here 4q-factorization predicts testable relations
among isospin amplitudes in the large $N_c$ limit even for $X'\ne 0$. For $B\to
D X_u$ decays the four possible isospin amplitude are predicted in terms of two
factorized matrix elements, while for $B\to D\bar D X$ there are two non-trivial
relations among the seven isospin amplitudes. For these tests examples of
relevant decay modes include $B\to D^{(*)}\pi\pi$, $B\to D^{(*)} KK$, $B\to
D^{(*)}\bar D^{(*)}K$, etc. Further details may be found in Ref.~\cite{bgps}.

I would like to thank C.~Bauer, B.~Grinstein, and D.~Pirjol for collaboration on
results presented here.  This work was supported in part by the Department of
Energy under the grant DE-FG03-00-ER-41132.

\end{document}